\documentclass[twocolumn]{aastex631}
\usepackage[utf8]{inputenc}
\usepackage{mathtools}
\usepackage{comment}

\received{\today}
\revised{\today}
\accepted{\today}

\submitjournal{ApJ}

\shorttitle{Stellar Distributions in GC}
\shortauthors{Linial \& Sari}

\newcommand{\pfrac}[2]{\left( \frac{#1}{#2} \right)}
\newcommand{\pdiff}[2]{\frac{\partial #1}{\partial #2} }
\newcommand{\MBH}{M_{\bullet}}

\newcommand{\e}{\varepsilon}

\newcommand{\ttb}{\tau_{\rm 2B}}

\newcommand{\md}{m_{\rm d}}

\newcommand{\xmin}{x_{\rm min}}
\newcommand{\xmax}{x_{\rm max}}

\newcommand{\mmin}{m_{\rm min}}
\newcommand{\mmax}{m_{\rm max}}

\begin{document}
\title{Stellar Distributions Around a Supermassive Black Hole: Strong segregation regime revisited}

\correspondingauthor{Itai Linial}
\email{itai.linial@mail.huji.ac.il}

\author[0000-0002-8304-1988]{Itai Linial}
\affiliation{Racah Institute, Hebrew University of Jerusalem, Jerusalem, 91904, Israel}

\author[0000-0002-1084-3656]{Re'em Sari}
\affiliation{Racah Institute, Hebrew University of Jerusalem, Jerusalem, 91904, Israel}

\begin{abstract}
We present a new analytical solution to the steady-state distribution of stars close to a central supermassive black hole of mass $\MBH$ in the center of a galaxy. Assuming a continuous mass function of the form $dN/dm \propto m^{\gamma}$, stars with a specific orbital energy $x = G\MBH/r - v^2/2$ are scattered primarily by stars of mass $\md(x) \propto x^{-5/(4\gamma+10)}$ that dominate the scattering of both lighter and heavier species at that energy.
Stars of mass $\md(x)$ are exponentially rare at energies lower than $x$, and follow a density profile $n(x') \propto x'^{3/2}$ at energies $x' > x$. Our solution predicts a negligible flow of stars through energy space for all mass species, similarly to the conclusions of \cite{BW_77}, but in contrast to the assumptions of \cite{AH_09}. This is the first analytic solution which smoothly transitions between regimes where different stellar masses dominate the scattering.
\end{abstract}


\section{Introduction}
How are stars and stellar remnants distributed in the vicinity of a supermassive black hole (SMBH) in the center of a galaxy? \cite{Peebles_72} provided an early treatment of this question. He heuristically argued that the steady-state distribution of equal-mass particles around an SMBH will be characterized by a constant flux of stars through energy space, estimated as $\dot{N}(r) \approx N(r)/\ttb(r)$, where $N(r)$ is the local number of particles between $r/2$ and $r$, and $\ttb(r)$ is the local two-body relaxation time (e.g., \cite{Merritt_2013,Sari_Fragione_2019})
\begin{equation}
    \ttb(r) \approx \frac{P(r)}{N(r) \ln{\Lambda}} \pfrac{\MBH}{M_\star}^2\,,
\end{equation}
where $P(r)$ is the orbital period, $M_\star$ is the stellar mass and $\ln{\Lambda}$ is the Coulomb logarithm.
By setting $\dot{N}(r) = \rm const.$, he found a density profile with $n(r) \propto r^{-9/4}$.  \cite{BW_76} revisited this problem and calculated explicitly the net particle flux through energy space, by integrating over all weak scatterings between particles of a given energy and all other energy levels. They demonstrated that Peebles' constant-flux solution is non-physical, as it predicts a negative flux, i.e., a flow of stars away from the supermassive black hole to outer radii. Rather, \cite{BW_76}  demonstrated the existence of a \textit{zero-flux}, steady-state solution, yielding the famous density profile $n(r)\propto r^{-7/4}$, commonly referred to as the ``Bahcall-Wolf cusp". For this profile, drift due to dynamical friction is offset by stochastic two-body scatterings such that the net flux of particles in energy space vanishes. Since this cancellation cannot occur very close to the inner boundary (near the tidal radius, or the SMBH's horizon), the ``zero-flux" condition in fact implies a constant, negligible flux that is set by the inner boundary, such that $\dot{N} \approx N(r_{\rm in}) / \ttb(r_{\rm in}) \ll N(r)/\ttb(r)$ for $r \gg r_{\rm in}$. In other words, the particle flux through energy space prevailing in the Bahcall-Wolf solution is negligible with respect to the ``na\"ive" local flux, estimated as $N(r)/\ttb(r)$ for $r\gg\ r_{in}$.

In a subsequent work, \cite{BW_77} generalized their treatment to include multiple mass groups, assuming that the most massive stars are also most abundant. They found that the massive stars follow the single-mass Bahcall-Wolf profile, while lighter stars with $m<\mmax$ follow a somewhat shallower profile of $n_m(r) \propto r^{-(3/2 + m/4\mmax)}$. This solution, similarly to their single-mass case, satisfies zero particle flux for all mass groups simultaneously.

Unlike the regime considered by \cite{BW_77}, light stars typically dominate in numbers, especially towards the outskirts of the sphere of influence where more massive stars (or massive stellar remnants) are comparatively rare. 
The tendency towards energy equipartition in the interaction of a massive particle with a light particle causes the massive one to sink and segregate down the potential well, and it is therefore expected that these objects become locally more abundant at smaller radii.
Several authors have addressed this light star dominated regime, relevant to the outskirts of the sphere of influence, sometimes referred to as the ``strong mass segregation" regime. \cite{AH_09} have considered a two-population model, with abundant light particles and rare massive particles, of masses $m_L$ and $m_H$. They argued that in the limit that massive particles are sufficiently rare, i.e., $N_L \ll N_H (m_H/m_L)^2$, light particles settle onto the standard Bahcall-Wolf profile, while massive particles drift towards the central black hole via dynamical friction, following a constant-flux profile with $n_H(r) \propto r^{-11/4}$. \cite{Keshet_2009} then generalized these results to treat a continuous mass function and obtained analytical expressions for the stellar distribution functions. This work reproduces and generalizes the results of \cite{BW_77} for the massive-dominated case, reaffirming the conclusion that in this regime, the most massive particles settle onto the single mass Bahcall-Wolf solution with a density profile $n(r)\propto r^{-7/4}$, while lighter species follow a shallower density profile.

Over the past decade, several studies have addressed the expected distributions of stars in galactic nuclei numerically, using direct N-body calculations, Monte-Carlo simulations or by numerically integrating the Fokker-Planck equation \citep[e.g.,][]{BW_76,BW_77}. Such calculations have been used to constrain the rate and properties of various gravitational wave sources occurring in galactic nuclei \citep[e.g.,][]{Oleary_2009,Preto_Amaro_Seoane_2010,Merritt_2010,Amaro_Seoane_Preto_2011,Merritt_2011,Aharon_Perets_2016,Emami_2021}. Numerical calculations of stellar distributions have also been used in interpreting observations of the Milky Way's Galactic Center \citep[e.g.,][]{Vasiliev_2017,Baumgardt_2018,Generozov_2018,Taras_2019}, and in order to estimate the rates of tidal disruption events occurring in galactic nuclei \citep[e.g., ][and references within]{Taras_2019,Stone_2020}.

In this work, we find steady-state stellar distributions characterized by zero-flux, assuming a continuous mass function that scales as a power-law. The main goals of this paper is to generalize the Bahcall-Wolf solution to allow for a smooth transition from the regime where the stellar population is dominated by light particles on the outskits of the radius of influence to the regime where massive particles dominate, closer to the SMBH. 

We provide an explanation as to why zero-flux solution rather than constant flux solutions apply even in the light-dominated case. Our solution smoothly passes through regimes at which increasingly massive particles dominate the scattering, and provide a criterion for this transition between dominating mass groups.

Similarly to \cite{BW_76} and \cite{Keshet_2009}, we assume (1) spherical symmetry; (2) isotropic velocities; (3) Keplerian orbits (specifically, gravitational wave emission is neglected); (4) binarity is unimportant; (5) small, uncorrelated scatterings; (6) loss cone effects are neglected, (7) the inner and outer energy boundaries are separated by many of order of magnitude, $\xmin \ll \xmax$.

While this is essentially a mathematical paper, our aims stated in physical terms are to (a) clear out the question of zero-flux vs. constant flux solutions, where we show that practically, only zero-flux solutions are relevant, and (b) show how a system transitions from regions where heavy particles dominate the scatterings to regions at which light particles dominate. We show that when such a transition is continuous with a wide enough range of masses, the distribution of heavy mass objects falls exponentially with radius.

\section{Steady-state distributions}

\subsection{Energy-space Fokker-Planck Equation}
We characterize the stellar distributions through the function $f(x,m) = dN/(d^3 v \, d^3 r \, dm)$, encoding the phase space density per unit mass of particles of mass $m$ whose specific orbital energy is $x$. As common, $x$ is defined such that stars that are bound to the central black hole have $x>0$.
The distribution function $f(x,m)$ evolves according to the Fokker-Planck equation (e.g., \cite{BW_76}, \cite{AH_09}, \cite{Keshet_2009})
\begin{equation}
    \pdiff{f(x,m,\tau)}{\tau} = -x^{5/2} \pdiff{Q(x,m,\tau)}{x} \,,
\end{equation}
where $\tau$ is proportional to time and is appropriately normalized, while $Q(x,m)$ is the flux of stars of mass $m$ through energy $x$, given by
\begin{multline} \label{eq:Q_x_m}
    Q(x,m) = \int_{m_{\rm min}}^{m_{\rm max}} m' \, dm' \int_{x_{\rm min}}^{x_{\rm max}} dx' \, \left\{ \max(x,x') \right\}^{-3/2} \\
    \times \left\{ m f(x,m) \partial_{x'} f(x',m') - m' f(x',m') \partial_{x} f(x,m) \right\} \,.
\end{multline}

This expression, which appears in \cite{Keshet_2009} generalizes the case of a discrete mass function discussed by to the case of a continuous mass function.

\subsection{Zero flux versus constant flux}
Steady-state solutions to the Fokker-Planck equation are characterized by a constant particle flux, i.e., $\partial Q(x,m)/\partial x = 0$ for all $x$ and $m$. \cite{BW_76} have demonstrated that the solution for the single mass case is obtained by demanding an even stricter condition - that the particle flux is not only constant, but rather vanishes, $Q(x,m) = 0$. This condition yields the so-called \textit{zero-flux} or \textit{negligible-flow} solution. Through this realization, Bahcall \& Wolf corrected an erroneous solution by \cite{Peebles_72} that was derived by invoking a constant, non-zero particle flux. The single-mass, zero-flux solution predicts that the phase space density of stars scales as a power-law with energy, $f(x) \propto x^{1/4}$.

Following a similar line of reasoning, a zero-flux solution for the discrete, multi-mass case was obtained by \cite{BW_77}. This solution, valid for the ``weak-segregation" regime, i.e., where the most massive particles dominate the stellar population, satisfies $Q(x,m_i) = 0$ for all masses $m_i$ simultaneously. This solution was later generalized for continuous mass functions by \cite{Keshet_2009}. Both works predict that the most massive particles follow $f(x,\mmax) \propto x^{1/4}$, whereas lighter particles, $m<\mmax$ follow a shallower profile, $f(x,m) \propto x^{m/4\mmax}$.

The two-population, strong-segregation regime discussed by \cite{AH_09} invokes zero-flux for the light particles, i.e., $f_L(x) \propto x^{1/4}$, and a constant, non-zero flux for the heavy particles, with $f_H(x) \propto x^{5/4}$. This solution was motivated by the assumption that if the massive particles are sufficiently rare, they should have no dynamical impact on the distribution of the light particles, that self-scatter and settle onto the single-mass Bahcall-Wolf cusp. However, as $x$ increases, this solution predicts that the massive stars become increasingly more abundant, until they become the dominant source of scattering, of both light and massive particles. Therefore, deep inside the potential well, the massive particles are expected to follow the single-mass, steady-state, zero-flux profile with $f_H(x) \propto x^{1/4}$. This solution therefore cannot be connected with the non-zero, constant-flux profile discussed by \cite{AH_09}, thus invalidating the assumptions at the base of this solution. The only way to reconcile this contradiction is if all mass groups simultaneously satisfy the zero-flux condition globally.

We demonstrate the existence of a global zero-flux solution, in the presence of a continuous mass function, where massive particles are rare. We seek an analytical solution to the equation $Q(x,m) = 0$, in the limit $\xmin \ll x \ll \xmax$, and $\mmin \ll m \ll \mmax$. We further assume that the overall population of stars follows a mass distribution that scales as a power-law, with
\begin{equation} \label{eq:dNdm}
    \frac{dN(m)}{dm} \propto m^\gamma \,.
\end{equation}

\subsection{Zero flux implies $p \propto m$}
A powerful result regarding steady-state distributions in the multimass case can be obtained as a corollary to the zero-flux condition, $Q(x,m) = 0$. For every two mass groups, $m_1$ and $m_2$ and for every energy bin $x$, we show that
\begin{equation} \label{eq:p_ratio_m_ratio}
    p(x,m_1) = p(x,m_2) \times \frac{m_1}{m_2} \,,
\end{equation}
where $p(x,m) = \partial(\log f(x,m))/\partial(\log x)$. First identified by \cite{BW_77} in the case of a finite number of mass groups, this result was generalized for a continuous mass spectrum by \cite{Keshet_2009}. For completeness we repeat this derivation here. Assuming that $Q(x,m_1) = Q(x,m_2) = 0$ for every $x$, consider the following vanishing expression
\begin{equation}
    Q(x,m_1)\cdot f(x,m_2) \cdot m_2 - Q(x,m_2)\cdot f(x,m_1) \cdot m_1 = 0 \,.
\end{equation}
Substituting $\partial_x f(x,m_i) = p(x,m_i) f(x,m)/x$ for $i=1,2$ and $\partial_{x'} f(x',m') = p(x',m') f(x',m')/x'$, we obtain
\begin{multline}
    0 = \left[ m_1 p(x,m_2) - m_2 p(x,m_1) \right] \times \\
    \int_{\mmin}^{\mmax} (m')^2 dm' \int_{\xmin}^{\xmax} \left\{ \max(x,x') \right\}^{-3/2} f(x',m') \, dx' \,.
\end{multline}

Since both integrands are non-negative, the expression in the first brackets must equal zero, leading to the desired conclusion, i.e., equation \ref{eq:p_ratio_m_ratio}.

The generality of equation \ref{eq:p_ratio_m_ratio} deserves a simpler explanation. There are two mechanisms that govern the orbital evolution of a given particle. The first, is a stochastic scattering coming from the discreteness of all the other masses. Since it is stochastic, it could move the particle in and out in phase space at equal probability, and the net flux depends on the derivative of the distribution function and on the number of scatters and their masses, and not on the mass of the scattered object. The other term, which is a systematic drift term, is essentially dynamical friction - the tendency of a given body to generate an over-dense wake behind it. That wake is proportional to the mass of the relevant body. Since these two effects have to balance each other in a zero flux solution, we get that the logarithmic derivative of the phase space distribution is proportional to the mass of the object, hence equation \ref{eq:p_ratio_m_ratio}.

\subsection{Physical intuition}

Since we are interested in a continuum of particle masses, and since we anticipate that massive particles tend to sink towards the center, we imagine that at ever increasing energy $x$, a larger mass group $\md(x)$ dominates the scattering. 
We are postulating that given the power-law mass function (equation \ref{eq:dNdm}), $\md(x)$ will also scale as a (positive) power-law of $x$, i.e.,
 \begin{equation}
     \md(x) \propto x^r,
 \end{equation}
where $r$ is a yet-to-be determined positive constant.

We further anticipate that at energies smaller than $x$, the distribution of mass $\md(x)$ will be a very steep while at energies larger than $x$ it would be very shallow. At the energy $x$, the distribution of mass $\md(x)$ would have an order of unity logarithmic slope.

\subsection{Mathematical ansatz}

Combining our physical intuition above, with equation \ref{eq:p_ratio_m_ratio}, we are lead to postulate  the following ansatz
\begin{equation}
    \frac{x}{f(x,m)} \pdiff{f(x,m)}{x} = r A m x^{-r} \,,
\end{equation}
where is $A$ is an arbitrary constant. The appearance of $r$ in the coefficient on the right hand side is for later convenience.

The solution to this differential equation is of the form
\begin{equation} \label{eq:multiMassGenSol}
    f(x,m) = C(m) \cdot \exp{\left(-mA x^{-r} \right)} \,,
\end{equation}
where the integration constant $C(m)$ is an arbitrary function of $m$. We assume again a power-law, $C(m) \propto m^{\alpha}$, where $\alpha$ is an exponent to be determined later. Plugging the above solution in the zero-flux equation $Q(x,m) = 0$, we obtain
\begin{multline}
    0 = \int_{\xmin}^{\xmax} dx' \, \left\{ \max(x,x') \right\}^{-3/2} \left[ x'^{-(r+1)} - x^{-(r+1)} \right] \\
    \int_{\mmin}^{\mmax} (m')^{\alpha+2} \, \exp{\left( -A m' x'^{-r} \right)} \, dm' \,.
\end{multline}
Under the substitution $z = Am' x'^{-r}$, the mass  integral ($dm'$) reduces to
\begin{equation}
    \left( x'^{r} / A \right)^{\alpha+3} \, \int_{\mmin A x'^{-r}}^{\mmax A x'^{-r}} z^{\alpha+2} e^{-z} \, dz \,.
\end{equation}
Note that the variable $z$ is in fact $p(x',r')/r$, i.e., proportional to the logarithmic derivative of $f(x',m')$. In the limit $m_{\rm min} \to 0$ and $m_{\rm min} \to \infty$, and if $A > 0$, the $dz$ integral converges if $\alpha>-3$, and equals $\Gamma(\alpha+3)$ where $\Gamma(n)$ is the Gamma function. The original equation reduces to
\begin{multline}
    0 =
    \int_{x_{\rm min}}^{x_{\rm max}} dx' \, \left\{ \max(x,x') \right\}^{-3/2} \\ \left[ x'^{-(r+1)} - x^{-(r+1)} \right] x'^{r(\alpha+3)} \,. \\
\end{multline}
Separating the integral to two ranges, $(\xmin,x)$ and $(x,\xmax)$, we are left with
\begin{multline} \label{eq:Q_zero_integrals}
    0 = \int_{\xmin}^{x} dx' \, \left[ x'^{-(r+1)} - x^{-(r+1)} \right] x'^{r(\alpha+3)} x^{-3/2} \\ +
    \int_{x}^{\xmax} dx' \, \left[ x'^{-(r+1)} - x^{-(r+1)} \right] x'^{r(\alpha+3)-3/2} \,,
\end{multline}
which can be expanded to
\begin{multline}
    0 = x^{r(\alpha+2) - 3/2} \left\{ \frac{1}{r(\alpha+2)} \left[ 1 - \pfrac{\xmin}{x}^{r(\alpha+2)} \right] - \right. \\ 
    \left. \frac{1}{r(\alpha+3)+1} \left[ 1 - \pfrac{\xmin}{x}^{r(\alpha+3)+1} \right] + \right. \\ 
    \left. \frac{1}{r(\alpha+2)-3/2} \left[ \pfrac{\xmax}{x}^{r(\alpha+2)-3/2} - 1 \right] - \right. \\
    \left. \frac{1}{r(\alpha+3)-1/2} \left[ \pfrac{\xmax}{x}^{r(\alpha+3)-1/2} - 1 \right]\right\} \,.
\end{multline}
The terms containing $(\xmin/x)$ and $(\xmax/x)$ raised to different powers are small compared to 1 if the following inequalities are satisfied
\begin{equation} \label{eq:multiMassInequalities}
    0 < r(\alpha+2) < \frac{3}{2} \;, \qquad 0 < r(\alpha+3) + 1 < \frac{3}{2} \,.
\end{equation}
In this case, taking the limit $x/\xmin \to 0$ and $\xmax/x \to \infty$, the integro-differential equation we started with reduces to the following simple algebraic equation
\begin{multline} \label{eq:multiMassAlphaRAlgEq}
    0 =  \frac{1}{r(\alpha+2)} - \frac{1}{r(\alpha+3)+1} -  \\ \frac{1}{r(\alpha+2)-3/2} + \frac{1}{r(\alpha+3)-1/2}  \,.
\end{multline}

\subsection{Solutions}
The algebraic equation \ref{eq:multiMassAlphaRAlgEq} can be rewritten as
\begin{equation}
    0 = \frac{3(r+1)(4\alpha r + 10 r - 1)}{r(\alpha+2)(\alpha r + 3r + 1 ) (2\alpha r + 4r - 3)(2\alpha r + 6r - 1)}
\end{equation}

For $r\neq -1$, the following solution is obtained

\begin{equation} \label{eq:multiMassAlphaR_sol}
    \boxed{\alpha = \frac{1}{4r} - \frac{5}{2} \,.}
\end{equation}

The inequalities \ref{eq:multiMassInequalities} then correspond to the following limit on $r$
\begin{equation}
    -\frac{5}{2} < r < \frac{1}{2} \,.
\end{equation}
Combined with the requirement that the $dz$ integral is well-defined, we note that the relevant limits are in fact $0< r < 1/2$ and $\alpha > -2$.

\subsection{The mass function}

We now relate the value of $r$ to the overall abundance of particles of different masses, given in equation \ref{eq:dNdm}.
\begin{multline}
    \frac{dN}{dm} \propto \int_{\xmin}^{\xmax} x^{-5/2} f(x,m) \, dx \propto \\ m^\alpha \int_{\xmin}^{\xmax} x^{-5/2} \exp \left( -mA x^{-r} \right) \, dx \,,
\end{multline}
and by substituting again $z = A mx^{-r}$ the distribution integral reduces to
\begin{multline}
    \frac{dN}{dm} \propto m^{\alpha-3/2r} \int_{A m\xmax^{-r}}^{A m\xmin^{-r}} z^{(3/2r)-1} \exp \left( - z \right) \, dz \,,
\end{multline}
and in the limit $\xmin \to 0$ and $\xmax \to \infty$, the integral in the above expression converges to a constant ($=\Gamma(3/2r)$) if $r>0$, and after substituting the zero-flux condition (equation \ref{eq:multiMassAlphaR_sol}) we obtain
\begin{equation}
    \frac{dN}{dm} \propto m^{-\frac{5}{4} (2+1/r)} \,.
\end{equation}
Equating to $m^\gamma$, we have
\begin{equation}
    r = -\frac{5}{4\gamma + 10} \,,
\end{equation}
corresponding to
\begin{equation}
    \alpha = -\frac{\gamma + 15}{5} \,,
\end{equation}
and the full solution is given by
\begin{equation} \label{eq:multiMassSol_gamma}
    \boxed{f(x,m) \propto m^{-(\gamma+15)/5} \cdot \exp{\left(-m A x^{5/(4\gamma+10)} \right)} \,.}
\end{equation}

In this derivation, we assumed that $\alpha > -3$ and $r>0$. These assumptions are satisfied as long as $\gamma < -5/2$.

\subsection{Solution description}
For every particle mass $m$, the solution follows different qualitative regimes as a function of energy. Particles of this mass dominate the scatterings at an energy scale
\begin{equation}
    x_{\rm d}(m) \approx (mA)^{-(4\gamma +10)/5} \,.
\end{equation}
Around this energy scale, $p(x_{\rm d}(m),m) \sim 1$. At higher energies, $x \gg x_{\rm d}$, increasingly more massive particles become the dominant source of scatterings, the distribution flattens, with $p(x,m) \to 0$. Asymptotically, the phase space density for a given mass $m$ becomes constant at high energies, with $f(x \gg x_{\rm d} , m) \propto m^{-(\gamma+15)/4}$.
At lower energies   , $x \ll x_{\rm d}$, lighter particles dominate, and the distribution declines exponentially with decreasing energy. The logarithmic derivative scales as a power law of energy, given by
\begin{equation}
    p(x,m) = m \pfrac{-5}{4\gamma + 10} A x^{5/(4\gamma+10)} \,, \quad x \ll x_{\rm d} \,,
\end{equation}
with $5/(4\gamma + 10) = -r < 0$.

\subsection{Solution validity range} \label{sec:validity}
The inequalities in equation \ref{eq:multiMassInequalities} imposed by the convergence of the flux integral correspond to limits on the overall mass function, such that our solution is valid for $\gamma<-5$, ensuring that the dominating mass $\md(x)$ is an increasing function of $x$.

Outside this range, the integrals of equation \ref{eq:Q_zero_integrals} diverge at $\xmin \to 0$ and $\xmax \to \infty$, namely, due to interactions which are non local in energy space.

Non-local interactions are mediated through stars on highly eccentric orbits. The Fokker-Planck equation and the fluxes calculated above were derived under the assumption that the orbital eccentricities follow a thermal distribution. This assumption is represented in equation \ref{eq:Q_x_m} where the $dx'$ integral accounts for interactions between star of energy $x$ with stars of energies $\xmin < x' < \xmax$.

We speculate that in reality, these flux divergences due to scatterings from $\xmin$ and $\xmax$ will disappear, as stars on highly eccentric orbits will be scattered away, 
and they would not be able to be replenished on a short timescale, since the replenishment rate of the stars on these eccentric orbits is limited by the local relaxation time.

To conclude, we conjecture that the solution we find may be valid for mass function shallower than $dN/dm \propto m^{-5}$. However, a more quantitative treatment of the deviations from thermal eccentricity distribution and their impact on the apparent flux divergences, that is needed to support this conjecture, is beyond the scope of this work. 

\section{Dominating mass group} \label{sec:dominating_mass_group}
At every energy $x$, a certain mass group, $\md(x)$ dominates the scattering of all other mass groups.
When weighted appropriately, the distribution function is given by
\begin{multline}
    \int_{\mmin}^{\mmax} m^{3/2} f(x,m) \, dm \propto \\
    \int_{\mmin}^{\mmax} m^{\alpha + 3/2} \exp{\left(-mA x^{-r} \right)} \, dm \propto \\
    \propto x^{r(\alpha+5/2)}=x^{1/4} \,,
\end{multline}
where in the last step we used equation \ref{eq:multiMassAlphaR_sol} and we considered the limit $\mmin \to 0$ and $\mmax \to \infty$ such that the integral reduces to the Gamma function. Remarkably, this result is independent of $\gamma$.

\begin{figure}
    \centering
    \includegraphics[width=\columnwidth]{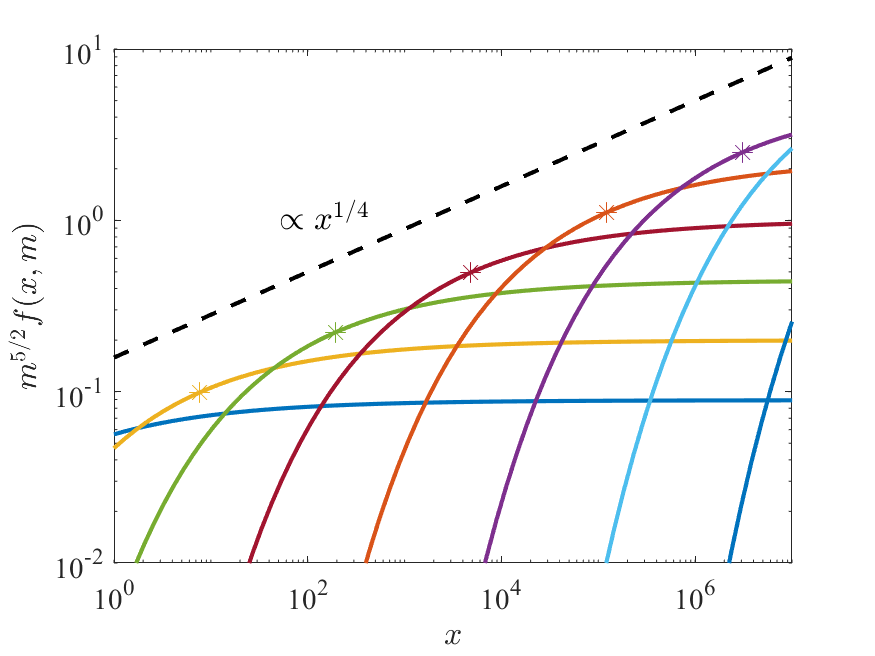}
    \caption{The phase-space density of stars of different mass groups, $f(x,m)$ for a mass function that scales as $dN/dm \propto m^{-6}$, as given in equation \ref{eq:multiMassSol_gamma}. The vertical axis, $m^{5/2} f(x,m)$ represents the phase space density of particles of mass of order $m$, multiplied by $m^{3/2}$. Different curves correspond to different mass groups, evenly spaced in logarithmic scale. Asterisks represent the energy at which each mass group locally dominates the scatterings. The dominating mass increases as the energy $x$ is increased.}
    \label{fig:MultiMass_multiplied}
\end{figure}

The expression $f(x,m) \, dm$ represents the number of particles per phase-space volume, $d^3 r \, d^3v$ with mass in the range $(m,m+dm)$. In the above integral, this density is weighted by $m^{3/2}$. Therefore, our result, encapsulated in equation \ref{eq:multiMassSol_gamma} can be stated in the following way: the ability to scatter is proportional $m^{3/2}$, and when this is taken into account, the distribution, multiplied by the $m^{3/2}$ factor, reproduces the single-mass Bahcall-Wolf profile. The factor $m^{3/2}$ can be heuristically understood as follows: the cross section for an object of mass $m$ to scatter a much lighter object scales as $m^2$, wheres its cross section in scattering much heavier objects is proportional to $m$. Since we are interested in scatterings of both lighter and heavier objects, the final result is expected to lie in between these two. Our analytical solution shows that it is exactly in the middle, with $m^{3/2}$.

In figure \ref{fig:MultiMass_multiplied} we show the distribution of stars of different mass groups, obtained for $\gamma = -6$. Every mass group was multiplied by $m^{5/2}$ (one extra power is required since $f(x,m)$ is the phase-space density per unit mass), such that the enveloping power law is the standard Bahcall-Wolf profile of $x^{1/4}$.

\section{Summary and discussion}
In this work we have presented a new analytical steady-state solution to the Fokker-Planck equation representing the flow of stars through energy space in the vicinity of a supermassive black hole, where the masses of the stars are assumed to follow a power-law mass function, $dN/dm \propto m^{\gamma}$.

Our solution builds upon the fact that as long $\mmin \ll \mmax$, and $\xmin \ll \xmax$, the problem becomes scale free in the range of intermediate energies and masses. Furthermore, we construct our solution by setting the particle flux of all mass groups through energy space to be zero. This construction is motivated by the zero-flux solutions of \cite{BW_76,BW_77}, and the understanding that the non-zero constant flux of heavy particles discussed by \cite{AH_09} cannot be maintained. With these assumptions, the initial integro-differential equation (eq. \ref{eq:Q_x_m}) can be greatly simplified into an algebraic equation (eq. \ref{eq:multiMassAlphaRAlgEq}) in $\alpha$ and $r$, the two parameters that appear in the general form of the solution (eq. \ref{eq:multiMassGenSol}). {\bf Thus we have obtained a fully analytic solution of the integro-differential equation; our analytic solution is given by the square box equation (\ref{eq:multiMassSol_gamma}).}

Our solution generalizes the works of \cite{BW_77} and  \cite{Keshet_2009}, in that we allow different mass groups to dominate the scatterings at different energy bins, rather than having only the most massive particles mass group dominate the scattering at all energies. As is illustrated in figure \ref{fig:MultiMass_multiplied}, in our solution, the dominating mass group gradually transitions from light to massive particles as the energy increases.

\cite{AH_09} coined the term ''strong-segregation" to describe the regime in which the lighter particles dominate the scattering, and therefore the density profile of heavy objects is very steep.
Many studies have reported the realization of the Alexander \& Hopman regime, through the identification of a steep density profile of the heavy species, $f_H(x) \propto x^{p_H}$, with $p_H \approx 0.5-1.5$ \citep[e.g.,][]{Oleary_2009,Keshet_2009,Preto_Amaro_Seoane_2010,Merritt_2010,Aharon_Perets_2016,Alexander_2017}.
While in their paper, strong segregation is maintained by a constant flux solution, we have shown that this is impossible, as it could not connect to a zero flux solution at higher energies where the heavy particles dominate.
In our interpretation, the reported steep density profiles are in fact the exponential rise that appears in our solution when a mass group transitions into dominating the scatterings (e.g., equation \ref{eq:multiMassSol_gamma} and figure \ref{fig:MultiMass_multiplied}). The key difference between the Alexander \& Hopman description and ours regards the flux of stars through energy space. Our claim is that stellar distributions settle onto profiles that yield a negligible flow of stars, much smaller than the local scattering rate, $N(r)/\ttb(r)$.

The solution we find is valid for sufficiently steep mass functions, with $\gamma < -5$. This constraints stems from the apparent divergence of the flux integrals due to non-local scatterings that occur for $\gamma > -5$. Nonetheless, we expect the solution be valid for somewhat shallower mass functions, since it is not clear that the eccentricities will continue to follow a thermal distribution once non-local interactions become important. We demonstrate that even without relaxing the assumption of thermal eccentricity distribution, the flux divergences are quite mild across a wide range of energies, as long as the mass function is not too shallow, i.e., $\gamma \gtrsim -5$. There are many uncertainties concerning the actual mass function of stars and stellar remnants within the radius of influence of a supermassive black hole. We provide a simple estimate of $\gamma$, in the limit of continuous star-formation, and a Salpeter IMF, $\xi(m) \propto m^{-2.35}$. For a wide range of stellar masses, the main-sequence luminosity scales as $L\propto m^{3.5}$, such that the main-sequence lifetime scales as $\tau_{\rm nuc} \propto m^{-2.5}$, and the present-day mass function therefore scales as $dN/dm \propto m^{-4.85}$.
This is just slightly out of the formal range of validity of our solution, and as we discussed in \S\ref{sec:validity}, our solution is likely to hold.

As a concrete example, we calculate the distribution $f(x,m)$ within the radius of influence of supermassive black hole of mass $\MBH$, as predicted by our model, assuming a mass function with $\gamma = -5$. We assume that stars of mass $\mmin$ are the dominant scatterers at the radius of influence, where the velocity dispersion is assumed to be $\sigma_h$. Considering the required normalization of the solution given in equation \ref{eq:multiMassSol_gamma} we obtain
\begin{multline}
    f(x,m) \approx 0.26 \times \\
    \frac{\sigma_h^3}{G^3 \MBH^2 \mmin^2} \pfrac{m}{\mmin}^{-2} \exp{\left(-\frac{m}{\mmin} \pfrac{x}{\sigma_h^2}^{-1/2} \right)} \,,
\end{multline}
where the prefactor satisfies the normalization condition, i.e., that the total enclosed mass equals $\MBH$
\begin{multline}
    \int_{\mmin}^{\mmax} dm \, m \int_{\sigma_h^2}^{\infty} dx \, f(x,m) \\ \int d^3 v \int d^3 r \, \delta( G\MBH/r - v^2/2 - x ) = \MBH \,,
\end{multline}
where $\xmax$ was taken to infinity.

Our solution naturally embodies the notion of mass segregation - particles of mass $m$ are exponentially rare at radii greater than $r_{\rm d}(m) \propto m^{-2}$ (for $\gamma = -5$), and then follow an approximate density profile of $n(r,m) \propto r^{-3/2}$ for $r < r_{\rm d}(m)$. In other words, heavy particles segregate to small radii around the supermassive black hole, while lighter particles are concentrated at larger radii.

Another important conclusion obtained from our solution concerns the transition of the dominating mass group, and the $m^{3/2}$ factor discussed in \S\ref{sec:dominating_mass_group}. This property can be stated as follows - for two mass groups, $m_L < m_H$, the transition between the region dominated by the scattering of one group to the other
occurs once their density ratio satisfies $n_L/n_H = (m_H/m_L)^{3/2}$. 

The flux equation we solve only accounts for weak, uncorrelated two-body scatterings. In particular, an important effect we neglect is the orbital dissipation due to the emission of gravitational waves. This effect  becomes important at energies of order $x_{\rm GW} \approx x_h^{1/3} x_s^{2/3}$ where $x_h$ and $x_s$ are the orbital energies at the radius of influence and close to the black hole's horizon \citep[e.g.,][]{Sari_Fragione_2019}).

\begin{acknowledgments}
The authors would like to thank Nicholas Stone for useful discussions and suggestions. This research was partially supported by an ISF grant. IL acknowledges support from the Adams Fellowship.
\end{acknowledgments}

\bibliography{Distributions}{}
\bibliographystyle{aasjournal}

\end{document}